\documentclass[preprint,prX,aps,showpacs,groupedaddress]{revtex4}

\usepackage[T1]{fontenc}
\usepackage[latin1]{inputenc}
\usepackage{latexsym}
\usepackage{array}
\usepackage{amsfonts}
\usepackage[centertags]{amsmath}
\usepackage{bm}
\usepackage{wasysym}
\usepackage[dvips,xdvi]{graphicx}
\usepackage{textcomp}
\usepackage{epsf}
\usepackage{hhline}
\usepackage{hyperref}

\newenvironment{eqnabc}{\begin{subequations}\begin{eqnarray}}{\end{eqnarray}\end{subequations}}

\begin{document}

\title{Revealing the pure confinement effect in glass-forming liquids\\by dynamic mechanical analysis}

\author{J.~Koppensteiner}\email{johannes.koppensteiner@univie.ac.at}

\author{W.~Schranz}\email{wilfried.schranz@univie.ac.at}

\affiliation{Faculty of Physics, University of Vienna,
Boltzmanngasse\,5, A-1090 Vienna, Austria}

\author{M.A.~Carpenter}

\affiliation{Department of Earth Sciences, University of Cambridge, Downing Street,
Cambridge CB2 3EQ, UK}
\date{\today}

\begin{abstract}
\noindent
The dynamic mechanical response of mesoporous silica with coated inner surfaces confining the glass forming liquid salol is measured as a function of temperature and frequency (1-100~Hz) for various pore sizes (2.4-7.5~nm).
Compared to former results on natural pores, a distinct acceleration of dynamics due to the removal of surface-related retardation of molecular dynamics is found now, which can be fitted by a homogeneous relaxation using an unmodified Vogel-Fulcher-Tammann relation.
This lubrication effect leads to a stronger decrease of the glass transition temperature $T_g$ with decreasing pore size.
The present data allow to quantify and separate competing side effects as surface bondings and negative pressure from the pure confinement induced acceleration of molecular dynamics with decreasing pore size. We analyze the dynamic elastic susceptibility data in terms of a recently proposed procedure [C.~Dalle-Ferrier et al., Phys.~Rev.~E \textbf{76}, 041510 (2007)], which relates the number $N_{corr}$ of molecules that are dynamically correlated to the three-point dynamic susceptibility. Assuming that $N_{corr}$ is representative of the size $\xi$ of a dynamically correlated region, i.e. $N_{corr}\propto \xi^3$ we find that $\xi$ grows when approaching the glass transition to about 3~nm, a size which was also reported previously using other techniques.
\end{abstract}

\maketitle

\section{Introduction}

Glass-forming materials have been produced by mankind for more than 6000 years. Despite several decades of intense research the transition of a liquid into its glassy state is still lacking a universal theory explaining both the increase of viscosity $\eta$ and molecular relaxation rates by 14 orders of magnitude \cite{cukiermann, ediger} without creating any long range order. A widely used explanation which goes back to Adam and Gibbs \cite{begriff_collectively_rearranging} is based on the assumption of cooperative rearrangement of molecules ("cooperative rearranging regions, CRR"), forming compact clusters of a typical size $\xi$. Such a subsystem of molecules can rearrange into another configuration independently of its environment.\\

The size of these groups of molecules is considered to grow to some nm as $T_g$ is approached \cite{hempel_xiauscp}. E.g. random first order transition theory of glasses \cite{Kirkpatrick1985, Lub} predicts $\xi=r_0\,0.51\,(ln\frac{\tau}{\tau_0})^\frac{2}{3}$. At $T_g$ where $\tau \approx 100$~s one obtains for typical values of $\tau_0 \approx 10^{-12}$~s and $r_0\approx 1$~nm, $\xi(T_g) \approx 5$ nm. Measuring such dynamic heterogeneities is one of the most important but at the same time difficult issues in the field of glass formation. Several experimental setups like specific heat spectroscopy \cite{hempel_xiauscp, Donthhuth}, multidimensional NMR \cite{Tracht, Reinsberg, Qiu} and non-linear dynamic susceptibility measurements \cite{Berthier}, etc. have been used to determine a possible growing length scale accompanying the glass transition. All these results agree in the fact, that the obtained cooperative regions are of the order of several nm near $T_g$, and are displaying a weak temperature dependence \cite{Erwin2002}. Recent computer simulations also confirm a rather modest growth of the dynamically correlated regions when approaching the glass transition from above \cite{scheidler3}. Very recently in a groundbreaking work Biroli~et~al. \cite{Biroli} found direct evidence for a growing dynamical length in supercooled liquids by applying inhomogeneous mode-coupling theory. Based on this theory these authors have recently developed a method to quantify the size of the correlated regions \cite{Berthier} by analyzing three-point dynamic susceptibilites. For a large number of supercooled liquids they could indeed confirm growing of the correlated regions when approaching the glass transition \cite{Ferrier2007}.\\

However, direct measurements of $\xi$ are extremely difficult, sometimes even impossible and one has to resort to indirect investigations. A widely used approach is confining the glass-forming liquid spatially, either in the form of thin films or by using mesoporous host matrices for confinement. If the size $d$ of the confinement is finite the cooperatively rearranging regions cannot grow beyond any bound, becoming saturated at $\xi=d<\infty$. This should lead to a confinement induced acceleration of the dynamics resulting in a  downshift of the glass transition, and even impeding it \cite{calorimetric3} as $\xi > d$.\\

Since the pioneer work of Jackson and McKenna \cite{calorimetric2} in 1991, uncovering a reduction of the glass transition temperature $\Delta T_g \propto 1/d$ in confinement of size $d$, a variety of confinement geometries and experimental methods have been used. Both weak and strong glass forming liquids, showing strong and weak interaction with the 2D or 3D confinement media were investigated. Single, double, even multiple transitions have been observed. Extensive topical overviews are found in Refs.~\onlinecite{review_Alcoutlabi}, \onlinecite{Simionesco} and \onlinecite{dielectric_overview}.\\

The abundance and diversity of experimental findings shows that an accurate discussion of side effects in discussing results of glass forming liquids in confinement is essential. Negative pressure due to mismatching thermal expansion coefficients of liquid and confining matrix is such a side effect. It was discussed by various authors \cite{patkowski, enthalpy_recovery} and sometimes even made responsible for the whole downshift of $T_g$ in confinement \cite{patkowski}. Being true, it would disprove the idea of a growing length scale of cooperativity. In a former paper \cite{koppensteiner} the authors have determined negative pressure effects for salol in natural un-coated pores of size 7.5 to 2.6~nm from high resolution thermal expansion measurements. An upper bound for the contribution of negative pressure to the total downshift of $T_g$ of $\approx$ 30\% was found.\\

A second, and much larger effect on the glass transition of liquids in confinement arises from the interaction of the molecules with the huge inner surface of confining host matrices which can take values up to 600 m$^2$/g (see Tab.~\ref{tab:samples_data}). Confined liquids tend to form H-bonds with the hydrophilic pore surface, which leads to an immobile surface layer of molecules and a retarded relaxation behavior at the glass transition.\\

In recent dynamic elastic measurements \cite{Schranz1, koppensteiner} of salol filled into matrices of Vycor and Gelsil with natural untreated pores we studied this competition between surface induced slowing down and confinement induced acceleration of the dynamics.
Here we present new results of salol confined in mesoporous Vycor and Gelsil with silanated pore surfaces. The results clearly show that silanation removes the liquid-surface interaction, leading to an enhancement of the molecular dynamics in the pores, resulting in a stronger downshift of $T_g$ as compared to the uncoated pores. These findings allow to separate the surface effect from confinement induced acceleration, and a simultaneous quantitative statement about negative pressure within one and the same measurement technique and confinement geometry. A comparison with previous dielectric spectroscopy data \cite{dielectric2, bulk_relax_arndt, Rittig} of salol in mesoporous matrices yields excellent agreement.\\

Using the recently proposed method of Berthier et al. \cite{Berthier} we determine the number $N_corr$ of dynamically correlated molecules as a function of temperature and pore size.

\section{Experimental}

\subsection{Sample preparation}

Vycor by Corning Inc., NY is produced via phase separation within a Na$_2$O-B$_2$O$_3$-SiO$_2$ melt, and subsequent acid leaching \cite{elmer}, which leaves a 98\% pure SiO$_2$ skeleton containing interconnected pores of random length, direction and density. A very narrow pore size distribution and an average ratio of pore length $l$ over pore diameter $d$ of $l/d \approx$4.35 is reported \cite{levitz}. Gelsil samples result from a sol-gel process \cite{nogues} and consist of randomly formed pure fused SiO$_2$ monodisperse spheres \cite{eschricht}, touching and penetrating each other, resulting in a mesoporous structure with a rather broad distribution of pore diameters, as N$_2$-sorption results showed. Samples are cut and sanded to gain the required orthogonal shapes, and cleaned in a 30\% H$_2$O$_2$ solution at 363 K for 24h. Drying is done under high vacuum ($10^{-6}$ mbar) at 393 K for another 24h.\\

In order to deactivate inner surfaces, OH$^-$ groups are replaced by OSi(CH$_3$)$_3$ trimethylsilyl groups via exposing cleaned samples to gaseous hexamethyldisilazane (HDMS, from Sigma Aldrich, purity 99.9\%) in a closed vessel at 330 K for 24h. Afterwards samples are again evacuated for 24h at 300 K. In order to check pore geometry and pore radius after silanation, mesoporous samples have been tested via N$_2$-sorption and BET/BJH analysis of the individual desorption isotherms \cite{rouguerol} experiments before and after HDMS treatment. Results are shown in Tab.~\ref{tab:samples_data}. Whereas adsorption data on Vycor does not show a significant change of pore geometry, in Gelsil samples surface area and pore volume were found to decrease strongly due to silanation. The effect of silanation on the pore diameter is $\approx$ 0.2 nm, corresponding to a HDMS layer thickness of $\approx$ 0.1 nm. For comparison Kremer et al. \cite{dielectric2} estimated the thickness of the silan layer from analysis of dielectric strength data as 0.38~nm.\\

Filling with the fragile low molecular weight glass former \cite{trofymluk} phenylsalicylate [salol, C$_{13}$H$_{10}$O$_3$, T$_m$(bulk)= 316 K]  is done by capillary wetting at 333 K for 12h. Filling fractions $f$ are evaluated via weighing clean and filled samples.\\

Some characterizing parameters for the glass-freezing behavior of bulk salol are: fragility index m - which usually varies from m=17 for strong glass formers to m=150 for fragile ones - takes for salol \cite{fragility} the value m=73, T$_g$(bulk)= 220~K (glass transition temperature defined at $\tau$=100~s) and  T$_{VF}$=175~K \cite{Tg_def_Richert} (Vogel-Fulcher temperature). The volume of a salol molecule is estimated \cite{salolsize2} as 0.282~nm$^3$ corresponding to a mean diameter of about 0.8~nm.

\begin{table}[h!]
\caption{$\textrm{N}_2$ adsorption characteristics and elastic moduli of untreated and silanated porous silica samples.}
\label{tab:samples_data}
\begin{center}
\begin{tabular}{llll}
\qquad&\qquad Gelsil 2.6&\;\; Gelsil 5&\;\; Vycor\\
\hline \hline
untreated\\
\hline
av. pore diameter\;(nm)&\qquad 2.6 &\;\; 5.0 &\;\; 7.5 \\
surface area\;($\textrm{m}^2$/g)&\qquad 590 &\;\; 510 &\;\; 70\\
pore volume\;($\textrm{cm}^3$/g)&\qquad 0.38 &\;\; 0.68 &\;\; 0.21 \\
porosity $\phi$\;&\qquad 0.36 &\;\; 0.54 &\;\; 0.40\\
bulk mod. K (GPa)\;&\qquad 9.6&\;\; 3.9 &\;\; 8.1\\
shear mod. G (GPa)\;&\qquad 7.7&\;\; 3.3 &\;\; 6.7\\
Young´s mod. Y (GPa)\;&\qquad 18.2&\;\; 7.7 &\;\; 15.8\\
\hline
silanated\\
\hline
av. pore diameter\;(nm)&\qquad 2.4 &\;\; 4.8 &\;\; 7.5 \\
surface area\;($\textrm{m}^2$/g)&\qquad 260 &\;\; 325 &\;\; 65\\
pore volume\;($\textrm{cm}^3$/g) &\qquad 0.15 &\;\; 0.4 &\;\; 0.19 \\
porosity $\phi$\;&\qquad 0.30 &\;\; 0.49 &\;\; 0.33\\
bulk mod. K (GPa)\;&\qquad 9.6&\;\; 3.3 &\;\; 9.1\\
shear mod. G (GPa)\;&\qquad 9.0&\;\; 3.9 &\;\; 9.0\\
Young´s mod. Y (GPa)\;&\qquad 20.6&\;\; 8.9 &\;\; 20.3\\
\hline \hline
\end{tabular}
\end{center}
\end{table}

\subsection{Dynamic mechanical analysis}

The dynamic mechanical response of samples of a typical size of 1\;x\;2\;x\;8 mm$^3$ in three point bending (3PB) mode yields the real and imaginary part of the complex Young's modulus $Y^*=Y'+i Y''$  within a frequency range of 0.01 to 100\,Hz applying static and dynamic forces up to 9.8 N. For further details see Refs.~\onlinecite{koppensteiner}, \onlinecite{Schranz1}, \onlinecite{dmaold1} and \onlinecite{dmaold2}. Temperature is controlled by gaseous nitrogen flow from 120 K to RT. DMA analyzers are decoupled from building vibrations, and electronics are shielded from a possible interference with the 50 Hz mains voltage frequency. The analyzers used are a series 7 DMA and a Diamond DMA, both built by Perkin Elmer Inc.

\subsection{Resonant ultrasound spectroscopy}

Due to contact errors a DMA experiment does not yield absolute values for elastic moduli. Therefore resonant ultrasound spectroscopy (RUS) was used to determine bulk and shear moduli of both natural and silanated mesoporous samples at room temperature. Orthogonal parallelepipeds of about 2.9 x 3.0 x 3.1 mm$^3$ were used to gather resonance spectra from 50 to 1100 kHz. For each sample 25 peaks resulting from excited resonant eigenmodes and corresponding overtones then were fitted via a Lagrangian minimization routine gaining bulk modulus K and shear modulus G (see Tab.~\ref{tab:samples_data}) with an accuracy of less than 1\%. For further experimental details see e.~g.~Ref.~\onlinecite{rus_details}. Young´s modulus Y was calculated from $Y=9KG/(3K+G)$ and used to calibrate DMA raw data at room temperature (see Figs.~\ref{fig:Gelsil5nm_sil_diamond_overview}, \ref{fig:data_vgl} and \ref{fig:pores_fit_overview}).

\section{Results and Discussion}

\begin{figure}
\begin{center}
\includegraphics[scale=0.75]{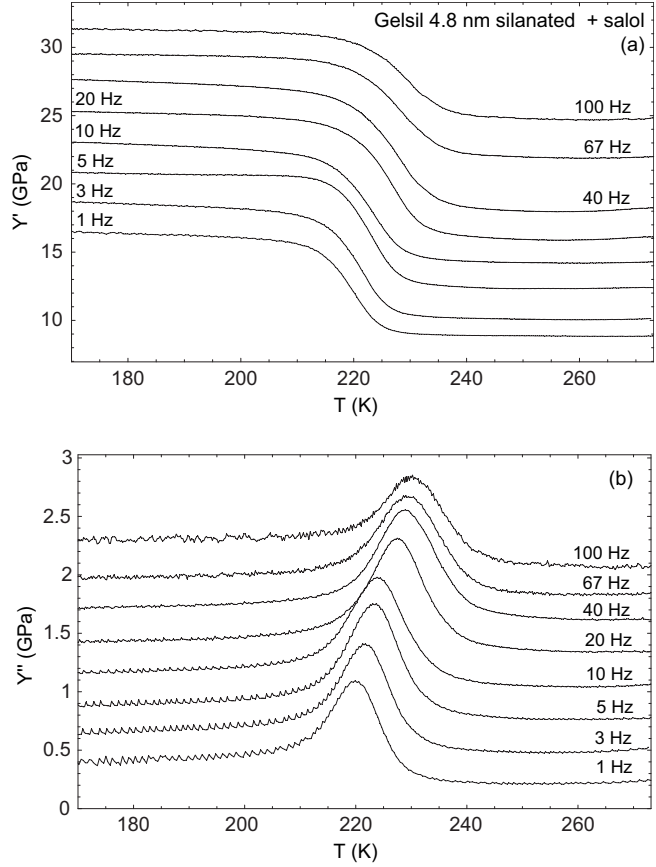}
\caption{Real (a) and imaginary (b) parts of the complex Young´s modulus of
silanated Gelsil (4.8~nm) filled with salol (filling fraction $f = 0.87$)
measured in three point bending geometry (Diamond DMA). 1Hz signal are original data, other signals are offset for sake of clarity.}\label{fig:Gelsil5nm_sil_diamond_overview}
\end{center}
\end{figure}

As an example, the dynamic elastic response (1\,Hz - 100\,Hz) of salol in Gelsil with silanated pores of 4.8~nm diameter is shown in Fig.~\ref{fig:Gelsil5nm_sil_diamond_overview}. The data for 2.4 and 7.5~nm pores look very similar and are hence not presented here.\\

\begin{figure*}
\begin{center}
\includegraphics[scale=0.72]{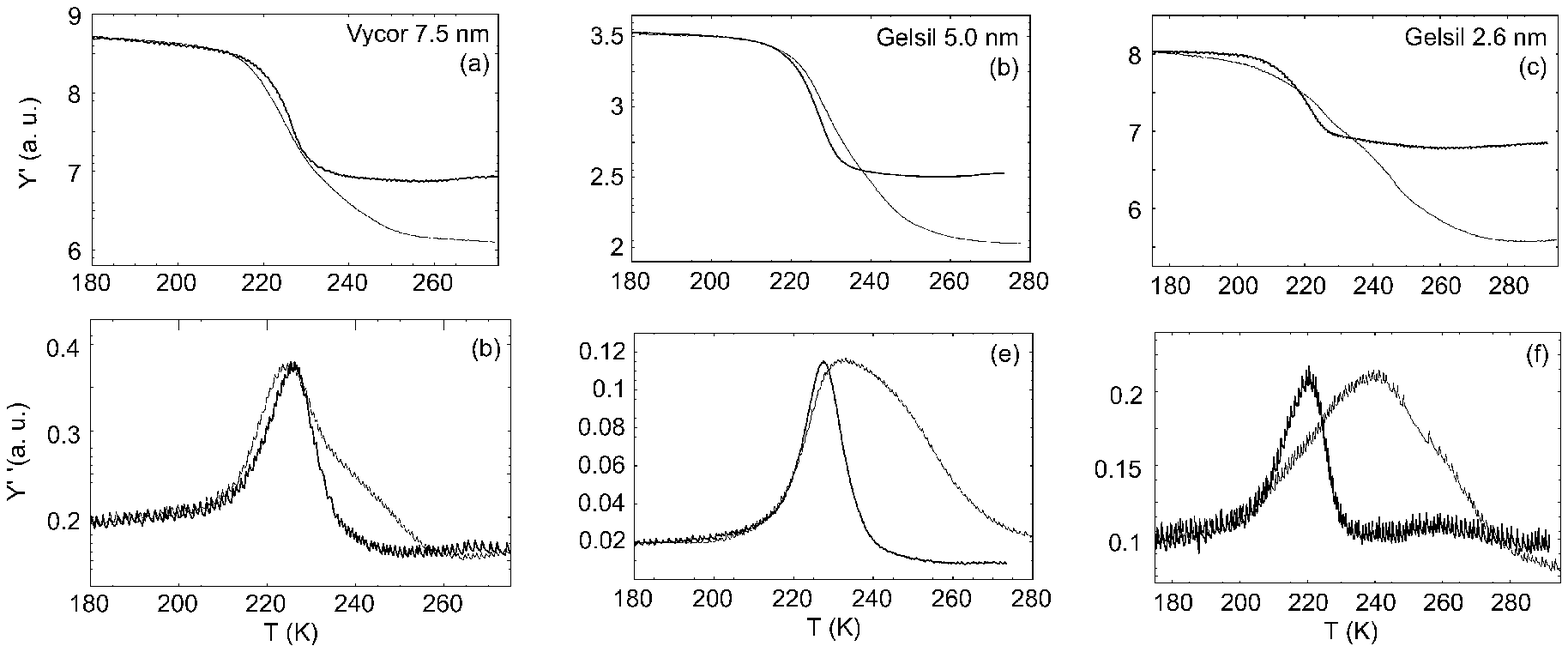}
\caption{Comparison of DMA data: thin lines are data for untreated inner surface, thick lines are new data for silanated pore surfaces. Data are presented in arbitrary units since in a DMA experiment contact losses do not allow to quantitatively compare $Y'$ and $Y''$ signals of two different samples.} \label{fig:data_vgl}
\end{center}
\end{figure*}

Fig.~\ref{fig:data_vgl} gives a comparison between the recent results of salol in natural untreated pores \cite{koppensteiner} and the new data on silanated samples. The most striking feature of the silanated samples is the absence of a double peak structure and the shoulder in $Y''(T)$ and $Y'(T)$, respectively. In untreated samples a significant part of the confined liquid sticks to the pore surface due to hydrogen bonding, thus being retarded in relaxation dynamics. This leads to a local dependance of relaxation times across the pore section (see Fig.~\ref{fig:pore_model}) causing an extra glass transition at higher temperature relative to the one of the core molecules. Assuming a spatial distribution of Vogel-Fulcher temperatures \cite{zorn1} $T_0(r)=T_{00}+k/(R-r+r_p)$, which translates via $\tau(r)=\tau_0 \cdot exp[E/(T-T_0(r))]$ to a distribution of relaxation times we were able to fit the stepwise change in $Y'(T)$ and the double-peak-structure in $Y''(T)$. A detailed analysis can be found in Ref.~\onlinecite{koppensteiner}.\\

For silanated surfaces we now find only bulk-like relaxation, i.e. just one step in $Y'(T)$ and one single narrow peak in $Y''(T)$. Peaks do shift with measurement frequency, as Fig.~\ref{fig:Gelsil5nm_sil_diamond_overview} shows. Since the liquid-surface interaction is removed now, we do not take into account any radial dependence of the relaxation time (see Fig.~\ref{fig:pore_model}), but use a single homogeneous Vogel-Fulcher-Tammann equation
\begin{equation}
\tau(T)=\tau_0\cdot exp \left[\frac{E}{T-T_0(d)}\right]
\label{eqn:VFT_temp}
\end{equation}
with the pre-exponential factor $\tau_0$, the activation energy $E$ and the Vogel-Fulcher temperature $T_0(d)$, depending on the pore diameter $d$.\\

\begin{figure*}
\begin{center}
\includegraphics[scale=0.8]{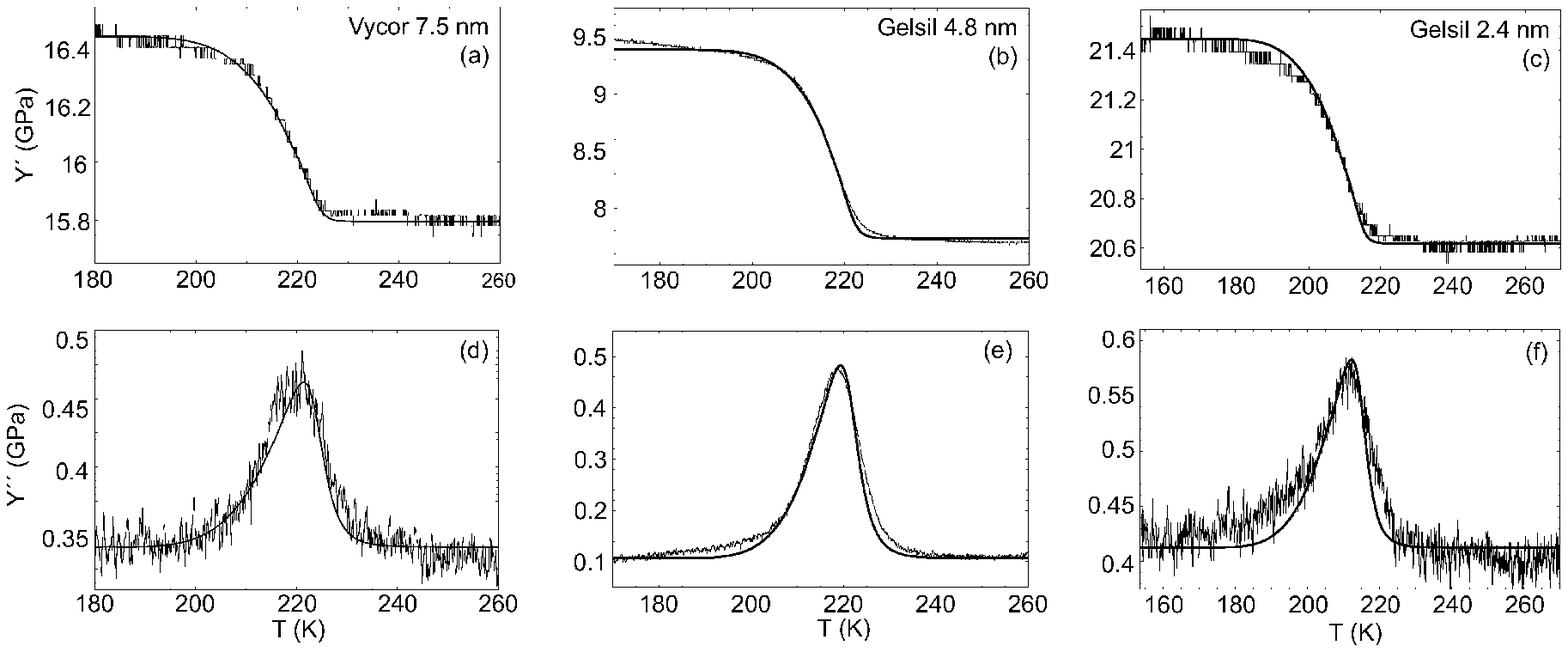}
\caption{Real part $Y'$ and imaginary part $Y''$ of different porous samples filled with salol. Lines are fits using Eqs.~(3a,b) and (\ref{eqn:VFT_temp}) with parameters of Tab.~\ref{tab:FitparameterVycorGelsil}.} \label{fig:pores_fit_overview}
\end{center}
\end{figure*}

\begin{figure}
\begin{center}
\includegraphics[scale=0.5]{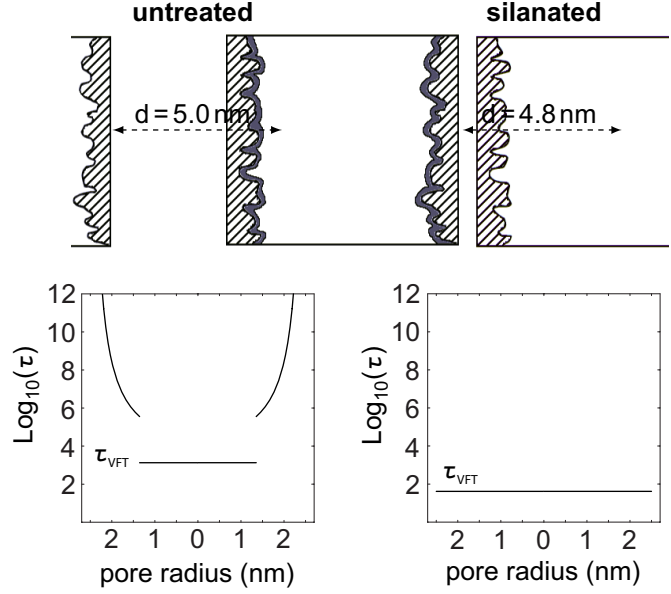}
\caption{Modeled relaxation time in untreated and silanated pores of Gelsil 5 from Eqn.~(\ref{eqn:VFT_temp}) used in Ref.~\onlinecite{koppensteiner} and in Eqs.~(3a,b) for fits of data in Fig.~\ref{fig:pores_fit_overview}b and e herein.} \label{fig:pore_model}
\end{center}
\end{figure}

\begin{figure}
\begin{center}
\includegraphics[scale=0.78]{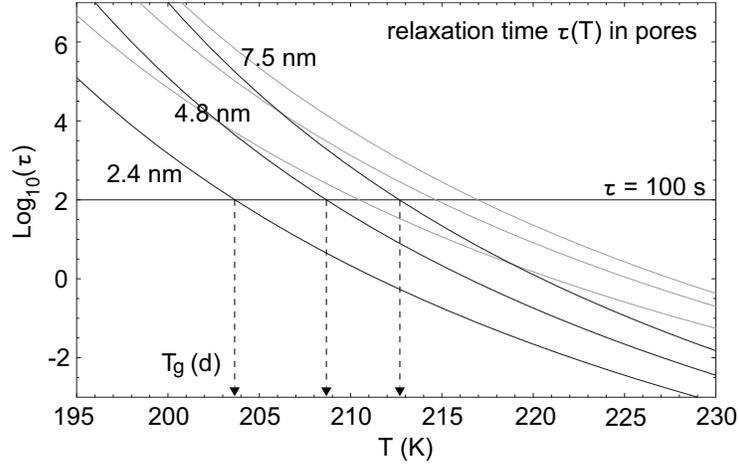}
\caption{Relaxation time in pore centers calculated from Eq.~(\ref{eqn:VFT_temp}) with corresponding parameters from Tab.~\ref{tab:FitparameterVycorGelsil}. Horizontal line shows $\tau=100$ s. Gray lines are relaxation times in untreated pores from Ref.~\onlinecite{koppensteiner}.} \label{fig:pore_center_tau}
\end{center}
\end{figure}

\begin{figure}
\begin{center}
\includegraphics[scale=0.85]{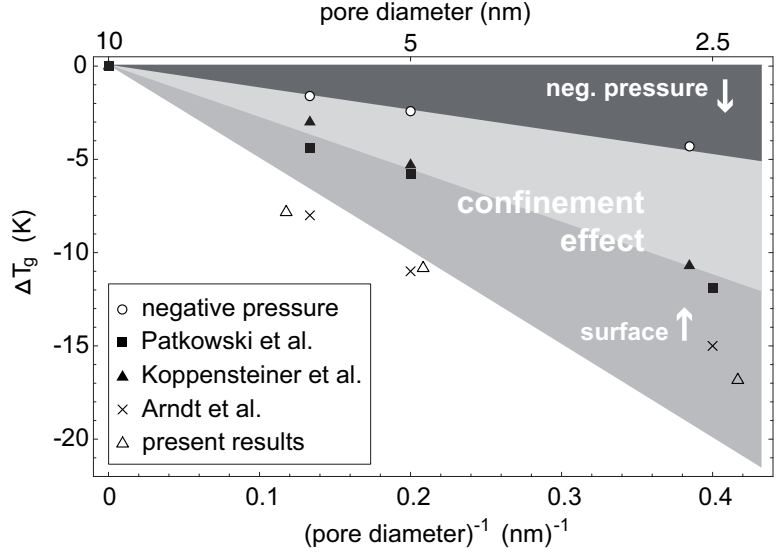}
\caption{Shift of glass transition temperature against $(\textrm{pore diameter})^{-1}$. Open circles display the maximum negative pressure contribution (see chapter III.~C.~of Ref.~\onlinecite{koppensteiner}), boxes are $\Delta T_gs$ from the same reference, filled triangles show literature values from Ref.~\onlinecite{patkowski}. Open triangles are the present results, and crosses mark corresponding literature data from Ref.~\onlinecite{dielectric2}.}
\label{fig:Tg_downshift_silanated}
\end{center}
\end{figure}

\begin{figure}
\begin{center}
\includegraphics[scale=1.1]{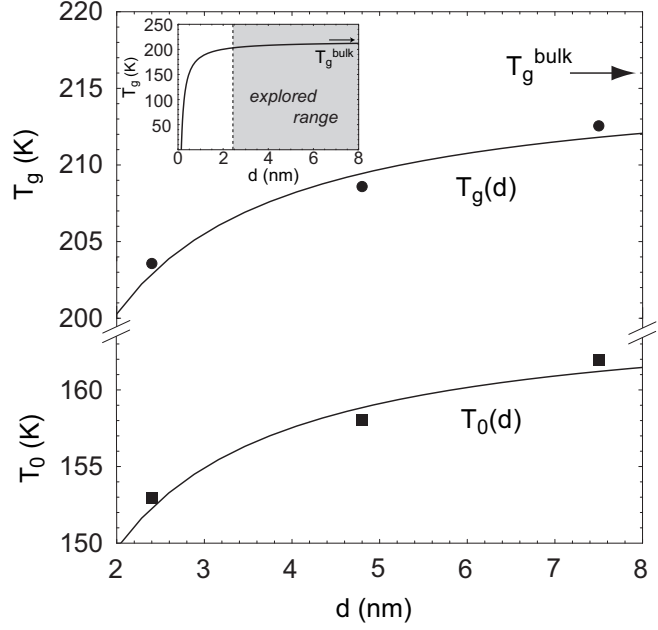}
\caption{Shift of glass transition temperature $T_g$ and Vogel-Fulcher temperature $T_0$ against pore diameter $d$. The points are determined from fitting the experimental data with Eqs.~(3a,b) and (\ref{eqn:VFT_temp}), where $T_g(d)$ is obtained from $\tau(T,d)=100$~s, as shown in Fig.~(\ref{fig:pore_center_tau}). The lines are fits using Eq.~(\ref{eqn:Tg(d)}) and the relation $T_g-T_0=\frac{E}{ln(100/\tau_0)}=50.6$~K, as indicated in the text.}
\label{fig:Tg_downshift_calculated}
\end{center}
\end{figure}

\begin{figure}
\includegraphics[scale=0.9]{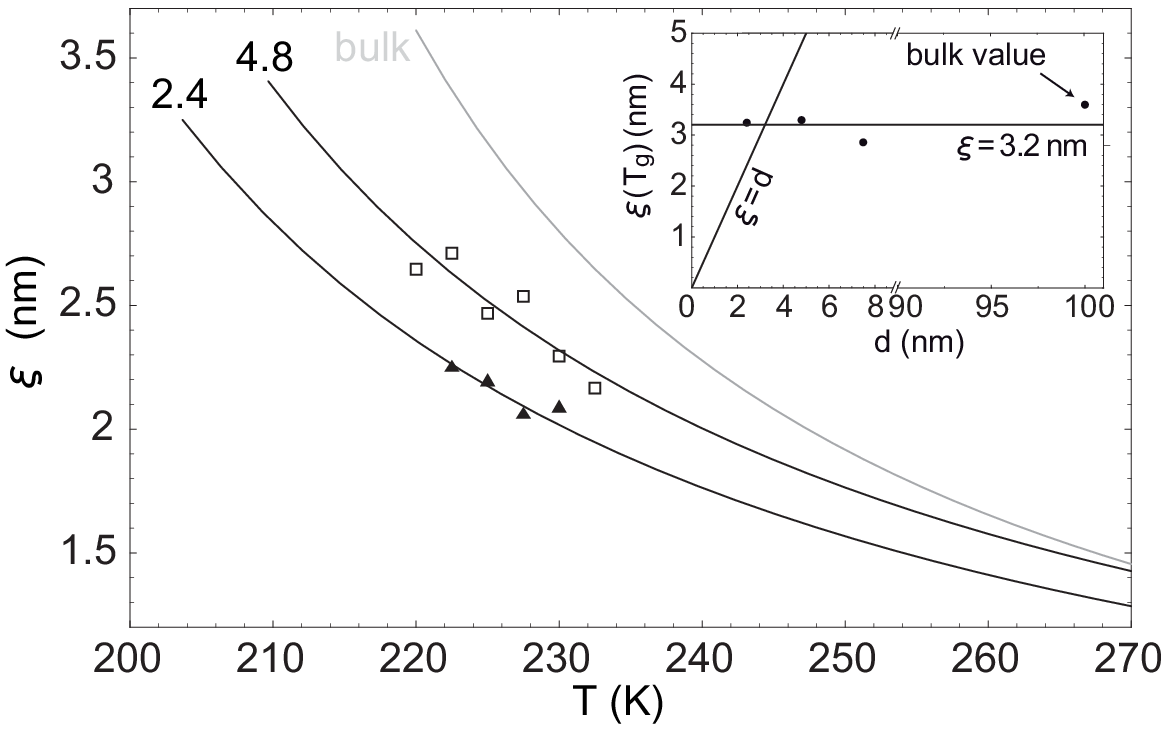}
\caption{Size $\xi$ of dynamically correlated regions. Lines are calculated from Eq.~(\ref{eqn:Ncorrfromfits}) using the procedure described in the text. Symbols are calculated from DMA data using Eq.~(\ref{eqn:Ncorr}). Inset shows $\xi$ at $T=T_g$ for various pore sizes.}
\label{fig:Ncorr}
\end{figure}

Similar as in our previous work \cite{koppensteiner, Schranz1} a Cole-Davidson relaxation is used to model dynamic mechanic susceptibility in terms of (now radially monodisperse) relaxors
\begin{equation}
Y^*(\omega) \propto \frac{1}{(1+i \omega \tau)^\gamma}
\label{eqn:cole_davidson}
\end{equation}
with $\omega=2 \pi \nu$, $\nu$ being the measurement frequency, and the broadening parameter \cite{footnote_gamma} $\gamma$.
Separating real and imaginary part of $Y^{*}=Y'+i Y''$ leads to
\begin{eqnabc}
Y'(T)=1-\Delta Y \cdot \frac{cos[\gamma\cdot arctan(\omega\tau(T))]}{[1+\omega^2\tau(T)^2]^\frac{\gamma}{2}}\\
Y''(T)=\Delta Y \cdot \frac{sin[\gamma\cdot arctan(\omega\tau(T))]}{[1+\omega^2\tau(T)^2]^\frac{\gamma}{2}}
\label{eqn:YstrichYzweistrichlang}
\end{eqnabc}
Eqs.~(3a,b) and Eq.~(\ref{eqn:VFT_temp}) are used to fit the data given in Figs.~\ref{fig:Gelsil5nm_sil_diamond_overview} and \ref{fig:data_vgl}. We point out that the pore radius $R = d / 2$ enters as a fitparameter in the pore size dependence of the Vogel-Fulcher temperature $T_0(d)$. Fits and corresponding parameters are shown in Fig.~\ref{fig:pores_fit_overview} and Tab.~\ref{tab:FitparameterVycorGelsil}, respectively.\\

Silanation  causes two main effects on the dynamics of the molecules within the pores:
1.) For untreated surface, we had to consider pore center relaxation times about two orders of magnitude higher \cite{koppensteiner} than obtained from the present analysis of silanated pores. Such an enhancement of mobility of molecules due to the absence of surface blocking in silanated pores was also observed in previous studies \cite{Rittig, Dosseh2006}. It probably reflects the fact, that the surface blocking of molecular mobility in uncoated pores slows down the dynamics of molecules also in the center of the pores.

2.) Present results and used fit parameters (Tab.~\ref{tab:FitparameterVycorGelsil}) show that the acceleration of the dynamics in silanated pores now leads to a much stronger downshift of $T_g$ with decreasing pore size as compared with the results of uncoated pores \cite{koppensteiner}. It reflects the pure confinement effect - i.e.~free of surface contributions - which leads to a downshift of $T_g$ due to limitation of the correlated regions by the pore diameter.\\

Unfortunately there is no unique theory which clearly relates these finite size effects to parameters characterizing the glass transition. However in recent computer simulations \cite{Varnik2002} of supercooled polymer films confined between two separated walls of distance $d$ it was shown that confinement leads to faster dynamics. The authors parametrized the size dependence of the relaxation time as $\tau(T,d) \propto exp\left[\frac{E(d)}{T-T_0(d)}\right]$ and found similar $d$-dependencies for the mode-coupling critical temperature $T_c(d)$ and the Vogel-Fulcher temperature $T_0(d)$ arguing, that $T_g(d)$ should also follow a similar curve (see Fig.~14 of Ref.~\onlinecite{Varnik2002}). As already mentioned above we here use the same dependence of the relaxation time (Eq.~(\ref{eqn:VFT_temp})) to fit our data, and obtain a very similar size dependence of $T_g(d)$ and $T_0(d)$ (Fig.~\ref{fig:Tg_downshift_calculated}) as Varnik et al.\\

A.~Hunt \cite{Hunt1994} has calculated finite size effects on the glass transition temperature in glass-forming liquids analytically using percolation theory, yielding
\begin{equation}
T_g(d) = T_g^{bulk}\left(1- c\frac{r_0}{d}\right)
\label{eqn:Tg(d)}
\end{equation}
where $r_0$ is the typical distance between molecules, which we approximate as the diameter of a salol molecule $r_0\approx0.8$ nm. Using a Gaussian distribution of energy barriers, the author obtained a value of $c\approx0.27$. A fit of our experimentally determined $T_g(d)$ (Table \ref{tab:FitparameterVycorGelsil}) with Eq.~(\ref{eqn:Tg(d)}) yields $c\approx0.13$ and $T_g^{bulk}\approx 216$~K (Fig.~\ref{fig:Tg_downshift_calculated}), which is in rather good agreement with the theoretical estimate. In this picture $T_g$ is reduced by confinement because the average barrier height for the molecules in pores is smaller than the so called "blocking" barrier, which is responsible for the glass freezing in the bulk.
We have also plotted the values of $T_0(d)$ of Table~\ref{tab:FitparameterVycorGelsil} in Fig.~\ref{fig:Tg_downshift_calculated}.
The corresponding line was drawn using the relation $T_g-T_0=\frac{E}{ln(100/\tau_0)}=50.6$~K, which is obtained from Eq.~(\ref{eqn:VFT_temp}) using the fitparameters $E=1750$~K and $\tau_0=10^{-13}$~s of Table~\ref{tab:FitparameterVycorGelsil}.\\

As already mentioned above, Dalle-Ferrier et al. \cite{Ferrier2007} have given an expression for the number $N_{corr,4}$ of molecules that are dynamically correlated over a time interval of the order of $\tau$ as
\begin{equation}
N_{corr,4}(T)=\frac{T^2}{\Delta C_p}\left(max_\omega\frac{\partial \chi(T,\omega)}{\partial T}\right)^2
\label{eqn:Ncorr}
\end{equation}
where $\Delta C_P$ in units of the gas constant R is the excess specific heat of the glass-forming liquid at constant pressure \cite{Ferrier2007} and $\chi(T,\omega)$ is a suitable dynamic correlation function. Very often glass-forming materials are studied by dielectric spectroscopy measurements and therefore the dynamic susceptibility is identified with the dielectric susceptibility. To estimate the number of dynamically correlated molecules, we apply two different procedures: In the first we are using $\chi(T,\omega):=\frac{Y'(\omega)-Y'(\infty)}{Y'(0)-Y'(\infty)}$ in Eq.~(\ref{eqn:Ncorr}) to analyze the data directly, i.~e.~ without any fit procedure in between. In the second case we rewrite Eq.~(\ref{eqn:Ncorr}) with Eq.~(\ref{eqn:cole_davidson}) yielding
\begin{equation}
N_{corr,4}(T)=\frac{T^2}{\Delta C_p}f(\gamma)^2 \left(\frac{\partial ln\tau}{\partial T}\right)^2
\label{eqn:Ncorrfromfits}
\end{equation}
where $\tau(T)$ is obtained from fits of the data in Fig.~\ref{fig:pores_fit_overview} and $f(\gamma)=\frac{sin[\gamma\, arctan(\frac{1}{\gamma})]+\frac{1}{\gamma}cos[\gamma\,arctan(\frac{1}{\gamma})]}{(1+\gamma^{-2})^{1+\gamma/2}}$ results from the Cole-Davidson dynamic response function and is the analog to the stretched exponential $\beta$ of the KWW-response function used e.g. in Eq.~(8) of Ref.~\onlinecite{Capaccioli2008}.
Fig.~\ref{fig:Ncorr} compares the temperature dependence of the dynamic correlation length $\xi$ calculated from the two different methods and applying the hypothesized relation $\xi=2\left(3N_{corr,4}V_{salol}/4\pi\right)^{1/3}$, which results in excellent agreement. We used DSC values \cite{trofymluk} of $\Delta C_p(d)$ reported for salol confined in mesoporous silica in a wide range of pore sizes from 2.6 to 26.4~nm. The inset of Fig.~\ref{fig:Ncorr} shows that the dynamic correlation length at the glass transition temperature $\xi(T_g)$ for various pore sizes determined from Eq.~(\ref{eqn:Ncorrfromfits}) is rather constant with a mean value of about 3~nm. This overall behaviour of $\xi(T_g,d)$ is very similar to calorimetrically determined characteristic lengths for salol in confined geometries \cite{Donth2000}.

\begin{table}
\caption{Fit parameters used in Eqs.~(3a,b) for fits of Fig.~\ref{fig:pores_fit_overview}.}
\label{tab:FitparameterVycorGelsil}
\begin{center}
\begin{tabular}{llll}
\qquad&\qquad Vycor&\qquad Gelsil5&\qquad Gelsil2.6\\
\hline \hline
$E$\;(K)&\qquad$1750$&\qquad$1750$&\qquad$1750$\\
$T_{00}$\;(K)&\qquad$161.5$&\qquad$158.5$&\qquad$154.5$\\
$\tau_0$\;(s)&\qquad$10^{-13}$&\qquad$10^{-13}$&\qquad$10^{-13}$\\
$\gamma$&\qquad$0.19$&\qquad$0.25$&\qquad$0.17$\\
\hline \hline
\end{tabular}
\end{center}
\end{table}

\section{Conclusions}

Results of extensive dynamic mechanical measurements of the glass forming liquid salol confined in mesoporous silica with silanated pores are presented. It turns out, that silanation can completely remove the liquid-surface interactions. As a result the confinement induced acceleration of the dynamics can be measured purely, which leads now to a much stronger (as compared to uncoated pores) downshift of the glass transition temperature $T_g$ with decreasing pore size due to the hindering of cooperativity. Using the results of percolation theory \cite{Hunt1994} we have calculated the downshift of $T_g$ with decreasing pore size, which fits our data very well. In Ref.~\onlinecite{Hunt1994} it is also shown, that finite size effects are expected to set in when the pore size $d\approx 7\,r_0-10\,r_0$ yielding $4.8-8$ nm for salol with $r_0\approx 0.8$ nm. This is in very good agreement with our observations.\\

We have also analyzed our dynamic elastic data obtained for the different pore sizes in terms of a newly proposed theory \cite{Berthier} which relates the size of dynamically correlated regions to the temperature derivative of the dynamical two-point correlation function, which in our case can be identified with the dynamic susceptibility $Y(\omega,T)$. The results clearly show an increase of the dynamic correlation length $\xi$ with decreasing temperature approaching $\xi(T_g) \approx 3.2~$ nm.
This value is very similar to the one obtained previously \cite{hempel_xiauscp} from DSC and TMDSC data.
For smaller pore sizes $\xi(T,d)$ at a given temperature shifts to smaller values which is concomitant to the systematic decrease of the relaxation time $\tau(T,d)$ and the resulting downshift of $T_g(d)$.  However at the glass transition temperature the dynamic correlation length is almost independent of the pore size.\\

Unfortunately at present there is no unique theory that relates the relevant parameters controlling the confinement effects in glass-forming materials to experimental data. The reason for this is that the microscopic mechanism behind the glass transition is still not completely understood and more theoretical (e.g.~of the type presented in Ref.~\onlinecite{Krakoviack2005}) and experimental work is required to close the gap of knowledge and understand confinement effects in glass-forming liquids.

\vspace{0.8cm}

{\bf Acknowledgements}:  We thank Marie-Alexandra Neouze and the Institute of Materials Chemistry from the Vienna University of Technology for the N$_2$-characterization of our samples. We also thank Irena Dreven$\check{\textrm{s}}$ek-Olenik and Miha Devetak from the Jo$\check{\textrm{z}}$ef-Stefan-Institute in Ljubljana for help concerning silanation, which was done within the ÖAD-WTZ project Sl~19/2009. Support by the Austrian FWF (P19284-N20) and by the University of Vienna within the IC Experimental Materials Science ("Bulk Nanostructured Materials") is gratefully acknowledged.


\begin{thebibliography}{99}
\bibitem{footnote_gamma} Note that by mistake Eq.~(1) of Ref.~\onlinecite{koppensteiner} contains the exponent $\gamma/2$ instead of $\gamma$.
\end{thebibliography}
\end{document}